# Micromagnetic evaluation of the dissipated heat in cylindrical magnetic nanowires

J.A. Fernandez-Roldan[1,*], D. Serantes[2,3], R.P. Del Real[1], M. Vazquez[1] and O. Chubykalo-Fesenko[1]

[1] Instituto de Ciencia de Materiales de Madrid, ICMM, 28049, Madrid (Spain). *Email address: jangel.fernandez.roldan@csic.es

[2] University of York, Heslington, York, YO10 5DD (UK)

[3] Universidade de Santiago de Compostela, 15703, Santiago de Compostela (Spain)

**Abstract**

Magnetic nanowires (NW) are promising candidates for heat generation under AC-field application due to their large shape anisotropy. They may be used for catalysis, hyperthermia or water purification treatments. In the present work we theoretically evaluate the heat dissipated by a single magnetic nanowire, originated from the domain wall dynamics under the action of an AC-field. We compare the Permalloy NWs (which demagnetize via the transverse wall propagation) with the Co fcc NWs whose reversal mode is via a vortex domain wall. The average hysteresis loop areas –which are proportional to the Specific Absorption Rate (SAR)– as a function of the field frequency have a pronounced maximum in the range 200MHz-1GHz. This maximum frequency is smaller in Permalloy than in Co and depends on the nanowire length. A simple model related to the nucleation and propagation time and domain wall velocity (higher for the vortex than for the transverse domain wall) is proposed to explain the non-monotonic SAR dependence on the frequency.

**TEXT**

Magnetic nanowires (NWs) are well-known to be promising candidates for multiple novel applications in information technology, spintronics and the development of new sensors and magnetic recording media[1–3]. One of the promising application areas is biomedicine, where magnetic materials have attracted interest for their heating potential and diverse functional capabilities[4]. Particularly, magnetic NWs have been investigated for different bio-medical applications such as cell separation[5,6] or as magnetic resonance imaging contrast agent[7], drug delivery[8], amperometric sensors for the detection of multiple organic compounds[9] and magnetic hyperthermia[10,11], where their internalization by cancer cells has been recently reported[12] (The origin of the cell destruction by NWs has not been identified yet). Apart from heating resulting from magnetic dynamics (i.e. Néel mechanism)[10], physical rotation of NWs under AC-field (Brown mechanism)[11] can also produce a direct physical damage of cancerous cells. The increased heating by magnetic NWs as compared to other potential candidates as nanoparticles comes in the first place from their larger shape anisotropy. Anticipating our findings, the largest heating response of NWs by Néel mechanism would correspond to 200 MHz-1GHz frequencies. Recently, Ni NWs have been investigated experimentally for their large response by remote excitation under radio-frequency fields with these frequencies[10]. Functionalized with Au to decrease the toxicity, they have been shown to induce destruction of pancreatic tumours[13]. On the other hand, these frequencies are traditionally considered outside the safe limits for hyperthermia treatment.

Magnetic NWs, nanoparticles and nanorods have been also proposed for green technologies as water purification agents for their capacity for removing heavy metals and pollutants[9,14]. In this case a large permanent moment of NWs could be very beneficial for their orientation under external fields. Additionally, if the AC-field is applied, the local heating produced by NWs can be used to promote specific catalytic reactions for waste or dye removal. The above examples indicate the necessity to understand the origin of the heating by hysteresis process in NWs which should be investigated first from theoretical point of view in order to predict the maximum heating response. The heating by hysteresis losses depends on the particular reversal mechanism, which in nanowires is usually mediated by the nucleation and subsequent propagation of a domain wall. The domain wall type –either Transverse Domain Wall (TDW) or Vortex Domain Wall (VDW) – depends on the election of specific materials and geometry which are tuned during the synthesis process. Each domain wall type exhibits different dynamical properties and thus they are expected to exhibit differences on the released heat under AC-fields.

Additionally, the use of magnetic NWs for heating has several important advantages in comparison to magnetic nanoparticles. Firstly, most of magnetic NWs have a large net magnetic moment due to the high shape anisotropy and, thus they are in the single domain state at the remanence while nanoparticles of the same volume may be in the multidomain state. As a result, a large magnetic moment leads to large mobility under applied magnetic field but smaller sensitivity to thermal fluctuations. Hence, we can expect orientation of NWs in magnetic fields similarly to chains of nanoparticles[15]. Such orientation would induce a better heating response than randomly oriented magnetic nanoparticles[16]. Secondly, nanowires (e.g., those electrochemically synthesized inside alumina templates)[2] have smaller dispersion of their physical sizes such as diameters and shape as in the case of magnetics nanoparticles. Such dispersion of physical properties leads to undesirable effects of over or infra-heating[16]. Finally, and importantly, the magnetic anisotropy of NWs comes predominantly from their shape owing to their high length to diameter aspect ratio. For a typical diameter of around 30 nm, and above 200 nm in length, its magnetic anisotropy is practically independent on the NW length and thus no dispersion in this value is expected unlike the nanoparticle case[17]. As a drawback, the coercive field $H_c$ of magnetic nanowires is not defined by the corresponding anisotropy field but is lower than it due to the mechanism of nucleation and propagation of domain wall.

In the present work, we theoretically model the heat dissipated by magnetic nanowires originated from the domain wall dynamics under the action of the AC-field. The hysteresis loops of Permalloy (Py) and Cobalt (Co) fcc cylindrical NWs with 30 nm diameter and variable lengths are calculated by means of micromagnetic simulations varying the amplitude and the frequency of the applied field. The choice of materials corresponds to the possibility to change the reversal mode keeping the same geometry. Overall both, quasistatic and dynamic simulations, show that the Py NWs demagnetize via nucleation of a pair of TDWs at the ends of the nanowire and their propagation towards each other, while the Co fcc NW reversal mode is by a pair of VDWs as depicted in Fig. 1(a-b). The specific diameter of NWs was chosen to be 30 nm which is the minimum diameter for which experimentally the NW physical and geometrical properties are homogeneous, as prepared by the electro-deposition methods.

Quasistatic calculations show that the hysteresis loops are almost length independent for lengths over 120 nm. Hence, we investigate the AC-driven hysteresis loops for lengths between 120 nm and 960 nm. The material parameters considered in our calculations are summarised in Table I indicating also the corresponding reversal modes for the quasistatic situation. The damping constant was set to a conventional Py value 0.012 for both materials.

**TABLE I**. Material parameters[2,18] and the reversal mode of the static hysteresis processes in simulated nanowires. The (111) lattice cell direction is parallel to the nanowire with Euler angles (θ, ϕ, ψ) = (0.96, 0, 0.61) for Co[18].

| Material | $M_s$ (kA/m) | $A_{ex}$ (pJ/m) | $K_1$ (kJ/m$^3$) | Reversal |
|---|---|---|---|---|
| Py | 796 | 13.0 | 0 | TDW |
| Co fcc | 1401 | 10.8 | -75; (1,1,1) | VDW |

We model NWs hysteresis loops under an applied magnetic field, $H(t) = H_{max} \cos(2\pi f t)$ with frequency $f$ and amplitude $H_{max}$, with the aim to assess the released energy per period proportional to the averaged over many periods area of the hysteresis loop, $<A(f, L, H_{max})>$, where $L$ is the NW length. The characterizing parameter of the dissipation is the released heat per volume here called Specific Absorption Rate (SAR), and defined as SAR $= <A(f, L, H_{max})> f$. The hysteresis loops of a single NW and different lengths are calculated by means of the micromagnetic package OOMMF[19]. For the sake of simplicity we focus on the case of the magnetic AC field applied parallel to the NW axis (assuming that the NWs orient either specifically under the dc field during the water purification treatment or under the AC-field during the hyperthermia treatment as was reported for magnetic chains[15]). This requires fixing the amplitude to $\mu_o H_{max} = 500$ mT, higher than the saturation field $\approx 300$ mT in all cases to ensure that in the static limit the hysteresis cycle corresponds to major hysteresis loop for both compositions.

Calculations start with the NW in the saturated state along the field direction. The magnetisation is recorded for several periods of the oscillating field, typically around 100 but in some cases, up to 1000 for the highest frequencies in order to surpass the transients and to ensure reaching a stationary behaviour with stable averaged values. Example of the last 3 hysteresis loops for various frequencies and a NW with length 480 nm is presented in Fig. 1(c-d). We clearly observe that for both materials the dynamical loops are wider than those corresponding to static hysteresis. At low frequencies ($f < 300$ MHz for Permalloy, $f < 600$ MHz for Co) the hysteresis loops are broader than the quasistatic one and have a pronounced tilt at both sides which is interpreted as a retarded nucleation and propagation of the domain wall. At high frequencies the loops are not closed.

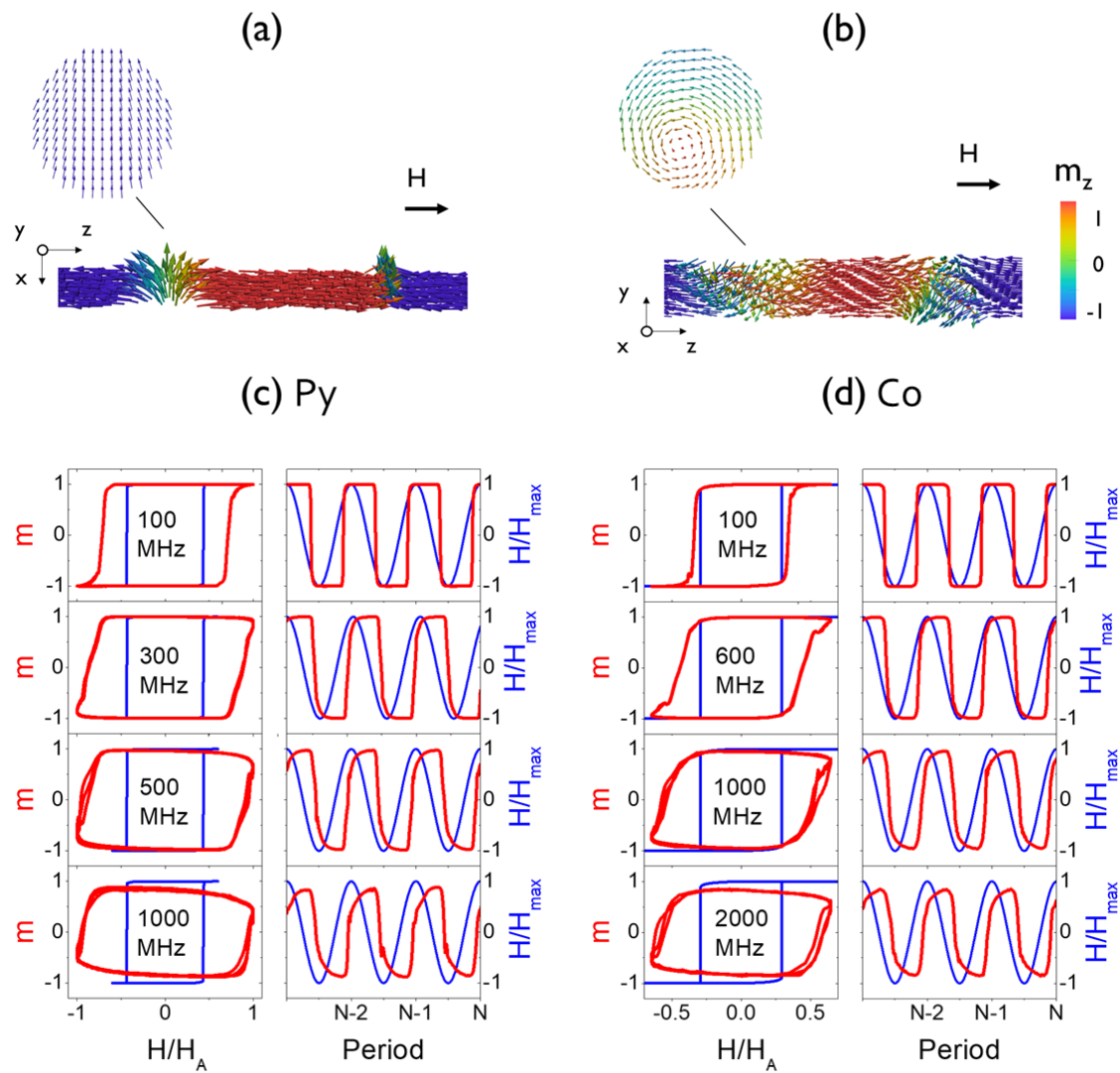

**FIG. 1.** (Colour online) (a) Tail-to-tail and head-to-head transverse domain walls propagation towards each other under an AC field. The left columns of (c) and (d) show the last three hysteresis loops (red curves) for 480 nm long nanowires of Permalloy and Co respectively for four selected frequencies. Quasistatic hysteresis loops are indicated by blue lines. The field axis is normalized to $H_A = 2K_{eff} / (\mu_o M_s)$ where $K_{eff} = |K_1| + \mu_o M_s^2/4$, i.e. 500 and 773 mT for Permalloy and Co, respectively. The right columns of (c) and (d) show the temporal evolution of the average magnetisation M (red) and H (blue) for the last 3 periods, normalized to the saturation magnetisation and field amplitude, respectively. N is the last period for each frequency.

Fig. 2 presents dynamical magnetisation configurations for Py NWs and various frequencies. At low and intermediate frequencies transverse domain walls nucleate at the ends of the nanowires, propagate towards each other and annihilate in the middle of the nanowire. During the propagation they are transformed into more complex structures and annihilate releasing energy by emission of spin waves along the entire length of the nanowire (See Fig. 2(a-b)). At high frequencies, the domain walls nucleate, propagate and annihilate in the same way but the relaxation to a saturated state by spin wave emission after the annihilation is not fully accomplished before the new reversal process starts as illustrated in Fig. 2(c-d). Finally, at very high frequencies, initial domain walls nucleate at the ends of the nanowire but they do not fully travel half of the length of the nanowire and neither annihilate during the field period. Simultaneously a new pair of domain walls is created at each end. As a result of the field periodicity, the accumulative effect is a train of distorted domain walls which looks like a multiple nucleation along the nanowire length as shown in Fig. 2e.

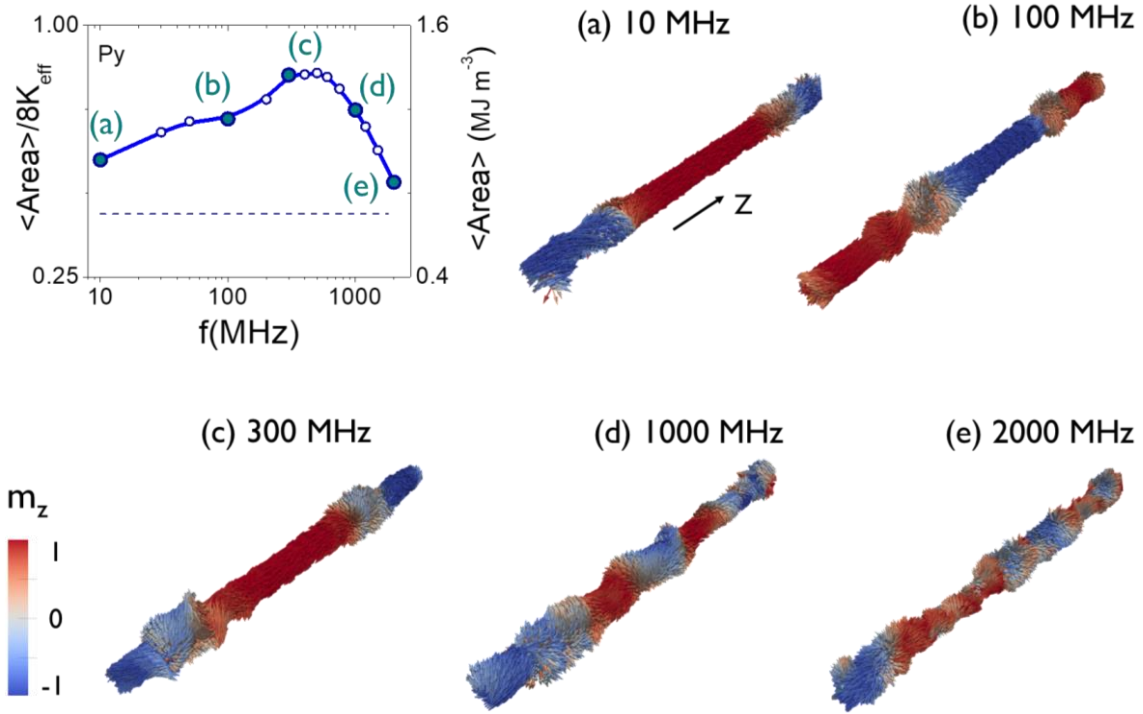

**FIG. 2**. (Colour online) (Top left) average hysteresis loop area for a permalloy 480 nm long nanowire. (a-e) Snapshots of the nanowire magnetisation configuration during the magnetisation reversal for the selected frequencies marked in the graph.

Furthermore, the shape of the domain wall is largely affected by frequency as well. Its shape becomes the more distorted, the higher the frequency is, compared with the quasistatic shape depicted in Fig. 1(a). This distortion is characterized by two effects (See Fig. 2(a-b)): First, a domain wall widening in length and second, a twist of the transverse component with the twisted angle increasing with frequency as shown in Fig. 2(b-c).

To evaluate the heating response, the hysteresis loop area for each period is individually integrated for every single loop and averaged. The mean value was obtained discarding a finite number of the first AC field periods. The standard deviations were evaluated for all the averaged quantities.

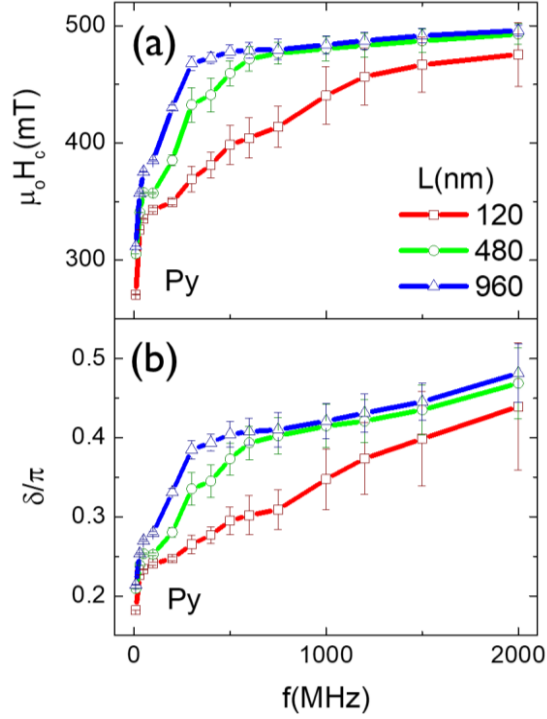

**FIG. 3.** (Colour online) The average coercive field $H_C$ (a) and the phase shift $\delta$ between the NW response and the applied field as a function of the AC-field frequency for three selected lengths of Permalloy nanowires.

First, our hysteresis loops in Fig. 1a show the frequency dependence of the coercive field for different lengths known as the dynamical coercivity (see Fig. 3 (a) for Permalloy, similar results are found for Co). The increase of the coercive field with frequency is well known (see e.g.[20,21] for the dots case) and appears due to the fact that the nucleation, depinning and propagation of the domain wall from NW ends requires certain finite time. The time is responsible for the decrease of the coercive field as a function of the nanowire length for fixed frequency. Indeed, in longer NWs domain walls require longer time to reach the centre of the nanowire at high frequencies which leads to the increase of the coercivity when the frequency increases. As a result, the coercive field of large NWs ($L \geq 480$nm) approach the $H_{max}$ at lower frequencies than the shorter NWs. Another prominent feature is the fact that Py nanowires saturate at lower frequencies than Co NWs. This is explained by larger mobility of vortex domain walls than of the transverse ones (see Ref.[22] and discussion below)

An additional distinctive feature of the hysteresis loops visible in Fig. 1 at high frequencies is the appearance of the phase shift ($\delta$) between the applied field and the magnetisation, which is related to the retarded nucleation and fixed propagation time, see Fig. 3 (b). The phase shift increases with frequency and reaches almost $\pi/2$ for large frequencies and is also dependent on the NW length.

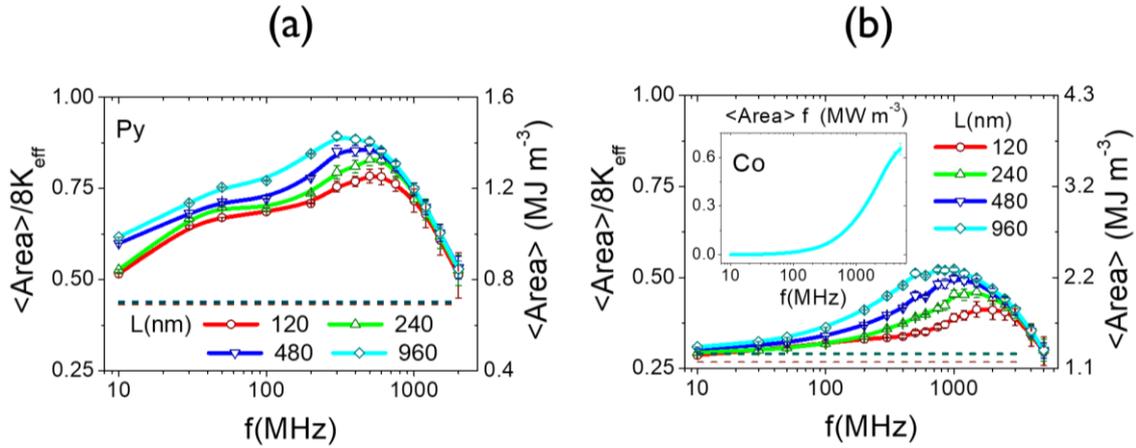

**FIG. 4.** (Colour online) (a) and (b) show the average hysteresis loop area $< A (f, L, H_{max})>$ as a function of the field frequency for different lengths normalized to 8 $K_{eff}$, for Py and Co, respectively. The inset of (b) is the Specific Absorption Rate as a function of frequency for the longest Co nanowire. The dashed line indicates the value extracted from the stationary hysteresis cycle.

The average area of the hysteresis loops as a function of the field frequency (Figures 4a and 4b) for the selected NW lengths is characterized by a maximum at a certain frequency, $f_{max}$. This frequency is larger for Co than for Py (see the comparison in Fig. 5b below). The maximum area values grow with the NW length. For the sake of comparison, the area in Fig. 4a, b is normalized to the maximum possible one, corresponding to the coherent rotation with the effective anisotropy of the long NW. In all cases hysteresis loops are wider than the static ones (see dashed line in Fig.4) as illustrated in Fig. 1. For frequencies $f < f_{max}$ the increase of the area is explained by the increase of the nucleation and coercive field. For frequencies over $f_{max}$ the NW does not reach saturation and the area decreases. Due to the dependence on the domain wall nucleation and propagation time, $f_{max}$ decreases with the nanowire length and is larger for Co (~ 1000 - 1500 MHz) than for Py (~ 300 - 500 MHz) reflecting again a larger velocity of the vortex domain wall in Co. The normalized hysteresis areas are larger in Py NWs (between 0.45 and 0.9 in the units of coherent rotation mechanism) than in Co (between 0.32 and 0.5) for the same frequency and length values due to the known fact that NWs with transverse domain wall have larger coercive field.

Finally, the specific absorption rate increases monotonically with f for both cases, being higher for Co NWs (with vortex domain wall).

To give a qualitative explanation for the frequency dependence of the hysteresis loop area we have developed a simple phenomenological model. First, we have evaluated domain wall mobility in Py and Co NWs by applying a constant magnetic field to a pre-existing domain wall and calculating the velocity of magnetisation change, considering that two domain walls are present. Our results show that the domain wall mobility has several qualitatively different regions of behaviour. In the field interval relevant to our modelling, the velocity is approximately linear with the applied field following the relation $v=v_0+\alpha (H-H_n)$, where $\alpha$ is the domain wall mobility and $H_n$ is the nucleation field for the stationary (major) hysteresis loop. For $H<H_n$ the domain wall mobility is different and is characterised by the final velocity $v_0$. Assuming that $H_n$ coincides with the coercive field of the stationary hysteresis loops for long NWs, i.e. $\mu_o H_n=215 mT$ for Py and $\mu_o H_n=255 mT$ for Co, the fitted values from the velocity dependence on field are the following: for Py $\alpha= 2600$ ms$^{-1}$/T, $v_0=1260$ m/s, and for Co: $\alpha=3000$ ms$^{-1}$/T, $v_0=2120$ m/s. Consequently, as we assumed previously, the vortex domain wall has larger mobility (in agreement with previously reported results in Ref.[22]) and larger velocities in the interval of interest.

Our simplified model assumes two equal rigid domain walls in the nanowire, where we assume that there is no magnetisation change for fields below the coercive one (squared major loop). In this case the domain wall propagation time is the main reason for the dynamical coercive field dependence on frequency or the nanowire length. Depending on the DW velocity and NW length, the DW could either propagate until reaching the other DW at the NW centre and thus annihilate there; or could be stopped by the AC field before approaching the other DW, and then to propagate in the opposite direction. The velocity can be integrated obtaining the DW position as a function of time, x(t), as:

$$x(t) = x(0) + (v_0 - \alpha H_N)t_N + \frac{\alpha H_{max}}{2\pi f}[\cos(2\pi f t_N) - \cos(2\pi f t)] \quad (1)$$

where $t_N$ is the nucleation time and x(0) is the position at that moment. For the sake of simplicity, we assume that $H_N$ is constant for all frequencies and that it takes the same value as the quasistatic coercive field ($H_C$). This assumption implies that $t_N=\mathrm{asin}(H_c/H_{max})/2\pi f$, and hence the equation (1) gives the magnetisation change as a function of frequency, field and nanowire length.

Even with these strong assumptions, the simplified model correctly reproduces the shape of the curves loops area versus the applied field frequency (see Fig. 5a for Co) as well as qualitatively several features: the decrease of the maximum frequency with the increase of NW length and the increase of the maximum value. As well as it qualitatively reproduces the fact that the hysteresis loop area is wider for Co as for Py due to larger domain wall velocities. The comparison with the direct simulations for the maximum frequency is presented in Fig. 5 b. The model works better for long nanowires since it assumes a rigid domain wall and neglects the time during which it is created from the nuclei at the nanowire ends. This process cannot be neglected for short nanowires. Additionally, the domain wall velocity was evaluated for long nanowires and the strength of dipolar field is different for the short ones. Furthermore, the agreement in Co is much better than in Py due to the fact that the hysteresis cycles are more squared. Finally, at high frequencies multiple domain walls are created which is also disregarded by the domain wall model.

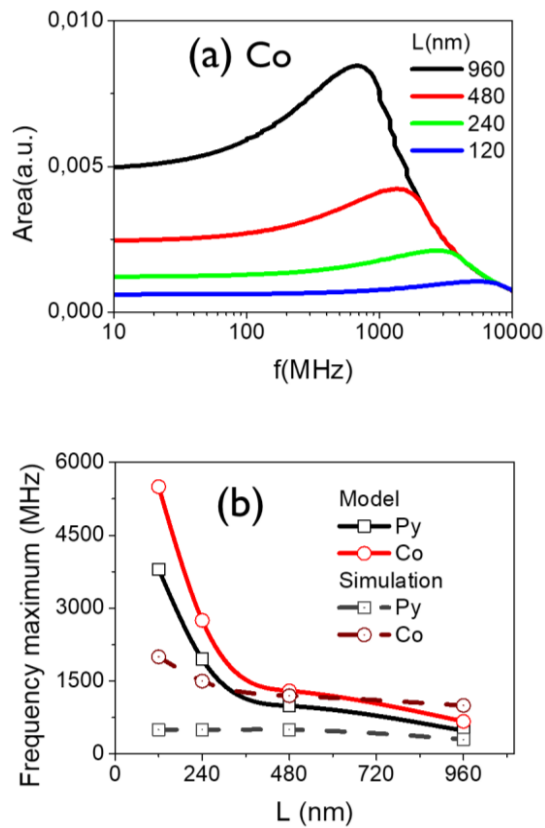

**FIG. 5**. (a) Area of the modelled hysteresis loop in Co nanowires as a function of frequency assuming the domain wall motion model (b) Comparison between the simulated and modelled values for the maximum frequency in Py and Co nanowires.

In conclusion, the heating response has a maximum at certain frequencies due to the influence of the nucleation and propagation time. This maximum is sensitive to the domain wall velocity, i.e. to the domain wall type and the NW length. Overall, Py NWs with transverse domain walls have larger dynamical coercive field which reaches almost 0.9 of the coherent rotation mechanism at frequencies of the order of 300-500 MHz. NWs with vortex domain walls (such as based on Co or Fe, i.e. with large saturation magnetization) have larger heating performance but the maximum appears at 1 GHz frequencies. The origin of the maximum in the heating performance is the balance between the propagation time and the NW length. At higher frequencies the domain wall does not have time to annihilate and several domain walls propagate. The difference between performance of Co and Ni lies in larger velocities of vortex domain walls as compared to the transverse one.

We believe that these findings show an important potential of nanowires for heating functionalisation in various applications such as the water purification treatment or magnetic hyperthermia. The tuneable heating response can be also used to promote certain catalytic reactions in desired applications.

**Acknowledgement**

This work has been supported by the Spanish Ministry of Economy, Industry and Competitiveness (MINECO) under the projects MAT2016-76824-C3-1-R and FIS 2016 – 78591-C3-3-R. D.S. acknowledges Xunta de Galicia for financial support (I2C Postdoctoral program). J.A. F.-R. acknowledges support from MINECO and European Social Fund though fellowship BES-2014-068789.